\documentclass[twocolumn,epjc3]{svjour3}  
\smartqed  
\RequirePackage{graphicx}

\journalname{Eur. Phys. J. A}
\begin{document}

\title{Mirror beta transitions} 

\author{
        K. Riisager\thanksref{e2,addr1}   }

\thankstext{e2}{e-mail: kvr@phys.au.dk}


\institute{Department of Physics and Astronomy, Aarhus University, Ny Munkegade 120, DK-8000 Aarhus C \label{addr1}
}

\date{Received: date / Accepted: date}

\maketitle

\begin{abstract}
Beta decays of mirror nuclei differ in Q-value, but will otherwise proceed with transitions of similar strength. The current status is reviewed: Fermi transitions are all very similar, whereas Gamow-Teller transitions can differ in strength by more than a factor two. The main cause of the asymmetries appears to be binding energy differences between the mirror systems.
\end{abstract}

\section{Introduction}
\label{intro}

The aim of the paper is to present an overview of the deviations from mirror symmetry in beta decays. More specifically beta decay transitions from mirror nuclei will be systematically compared. It should be noted that the term used in the title, ``mirror beta transition'', is also commonly used in a different context to designate beta transitions between $T=1/2$ mirror nuclei. Here the emphasis is on a comparison between two mirror decay processes, so only systems with $T \geq 1$ will be considered.

This is a subject that started more than half a century ago and much of the background for these investigations is given in the collection of chapters on isospin in nuclear physics edited by Wilkinson \cite{Wilk69}. A major difference from then to now is that many proton-rich nuclei are now accessible for detailed studies, recent reviews of nuclei close to the proton dripline and their decays can be found in \cite{Blan08,Pfut12,Pfut22}. The basic theory for beta-decay was already established then, but other ways of probing Gamow-Teller transitions have been explored since as described in \cite{Fuji11}, and symmetry tests in nuclear beta decay have of course evolved significantly, see \cite{Seve11,Vos15,Navi22} for recent review papers.

The current overview will start by a general look at mirror symmetry in nuclei in section \ref{aspect}, in particular the role of the Coulomb interaction. This is followed in section \ref{sec:beta} by brief general comments on the beta decay process and a discussion of two well-known relevant examples, namely that of superallowed Fermi transitions and of the strongly suppressed $^{14}$C decay. The following long section presents the current experimental situation on mirror beta decays. The results are then discussed briefly before section \ref{sec:final} concludes with a summary of the current state of the field.

\section{Aspects of mirror comparison}
\label{aspect}

We shall first discuss why a mirror symmetry in nuclei could be envisaged, in order to clarify how and when it is broken or (approximately) upheld.
This section describes several aspects of this question before turning to mirror beta transitions.

The proton and neutron masses differ by a fraction of $1.38 \times 10^{-3}$. The strong interactions between them are almost the same, but symmetry properties under particle exchange may limit which configurations of protons and neutrons are allowed. The isospin formalism is an elegant way of taking care of the symmetry properties, so one can discuss nuclear physics in terms of nucleons rather than neutrons and protons. However, protons and neutrons have different charges and magnetic moments and therefore different electromagnetic interactions.
Concerning the strong interaction, our current understanding of it is based on chiral effective field theory \cite{Mach11} and the availabality of many calculational frameworks \cite{Barr13,Carl15,Hamm20}. The small charge symmetry breaking (CSB) in strong interactions \cite{Mill90,Mill06} can be described both in meson ($\rho^0-\omega$ mixing) and in effective field theory frameworks and will generally give different contributions for mirror nuclei.

The weak interactions also distinguish between protons and neutrons, but in an isospin symmetric way in that the strength of the neutron-to-proton and proton-to-neutron transformations (the coupling to the $W$-bosons) are the same. However, this statement assumes second class currents \cite{Wilk00} to be negligible. These are small modifications to the weak currents, induced scalar for the vector part and induced tensor for the axial-vector part, that both will be proportional to the momentum exchange in the transition. They are expected to be induced, e.g.\ due to the symmetry breaking from the neutron-proton mass difference, but experimental limits are still above the expected level in the standard model.
See also \cite{Seve21} for a more complete theoretical overview and an in-depth discussion of the related (first class current) weak magnetism.

A crucial observation for the current discussion is that nuclei, in spite of having binding energies approaching 1\% of their rest mass, are loosely bound: separation energies of the least bound nucleons decrease rapidly when moving away from the line of beta-stability, so that far from all combinations of proton and neutron numbers will be particle bound; there are also quite few particle bound excited states in light nuclei. The quantum nature of nuclear systems is so pronounced \cite{Mott99} that wavefunctions of less bound nucleons will extend beyond the nuclear core. We can therefore expect wavefunctions in mirror nuclear systems to differ both due to differences in interactions, due to differences in their binding energy, and due to the small difference in proton and neutron mass that will affect the kinetic energies.

\subsection{The nuclear chart}
The Coulomb energy has a significant influence on the shape of the nuclear chart. It gives rise to the wellknown bending of the line of beta stability at higher masses, but it is striking that already two protons can bind six neutrons whereas two neutrons only bind two protons. The asymmetry in the masses of mirror nuclei is substantial and increases significantly with the mass number. The separation energies and Q-values for decay or reactions are differential in the mass surface and are therefore less affected, but rather quickly also become clearly asymmetric.

The proton and neutron driplines are therefore quite different. It is noteworthy that the proton dripline appears to be heavily influenced by shell-structure (and above oxygen also displays a clear odd-even effect in proton number). The double-magic $N=Z$ nuclei $^4$He,  $^{16}$O and $^{100}$Sn are the most proton-rich nuclei at their mass numbers, and $^{40}$Ca and $^{56}$Ni are only two nuclei from the dripline. At mass numbers in between, there is for particle-bound nuclei a substantial increase of proton number when passing neutron numbers 8 and 20. Most proton-rich nuclei therefore have the active protons and neutrons in the same major shell. For larger masses, the proton dripline slowly approaches the $N=Z$ line and crosses it at $^{100}$Sn. This of course limits the number of mirror (neutron-proton symmmetric) nuclei.
See also \cite{Pfut22} for a recent overview of the physics at the proton dripline.

\subsection{Binding energy and continuum effects}
Consider two mirror nuclei with their corresponding single-nucleon orbits. The Coulomb potential will give the major contribution to the difference in their binding energy, but
the difference in absolute binding energies of mirror nuclei cannot be explained with pure Coulomb contributions, this is known as the Okamoto-Nolen-Schiffer anomaly \cite{Shlo78}. Charge symmetry breaking strong interactions and proper account of nuclear radii appear to be able to explain the effect in modern shell-model calculations \cite{Mill06,Brow00,Mach01,Dufl02}.

The nucleon orbits will also be affected differently, have different separation energy and -- if situated close to the continuum -- have different spatial extension. The Coulomb potential will limit the effect on the otherwise systematically more loosely bound protons in the proton-rich nucleus, the angular momentum barrier will lessen differences for high $\ell$ orbits. In turn, the changes in the orbit structure may lead to a change in the structure of the states. When the energy differences (and even their ordering) of orbits in the two mirror systems differs, one can expect the occupations of individual orbits to differ as well. Changes may also arise from the different spatial structure of orbits in the mirror systems; here one can be quite sensitive to radial nodes in the wavefunction, but in the relevant region up to $^{100}$Sn only the second set of s and p orbits will have a radial node. As discussed further below, the excitation energies of states in mirror systems can also be expected to differ slightly.

In extreme cases, the binding energy effects or orbits close to the threshold can lead to the formation of halo systems \cite{Jens04,Tani13}, but the changes in structure can also persist to resonances in the continuum. 

\subsection{Isospin}
We shall employ the isospin sign convention where $T_z = (N-Z)/2$. The extent of isospin symmetry is best probed close to the $N=Z$ line \cite{Warn06}.

Beta decay is an obvious probe of isospin symmetry since the operators for nuclear beta decay transitions involve isospin, most clearly so for allowed Fermi transitions where the nuclear operator reduce to an isospin step-up or step-down operator. However, transitions with one-body operators, such as electromagnetic transitions, can also be divided into isoscalar and isovector parts and will also be useful for tests of isospin degrees of freedom \cite{Wilk69}.

In spite of the clear asymmety outlined above, isospin is surprisingly well conserved in nuclei. This is due \cite{Bohr69} to the fact that the Coulomb field varies rather slowly over the nuclear volume; for a uniform charge distribution the potential will be parabolic. Outside the nucleus Coulomb effects will be more pronounced. An instructive example is given by Garrido et al \cite{Garr08} that show how isospin mixing in the $^6$Li($2^+$) state peaks outside the nucleus at intermediate distances and decreases towards the asymptotic region.
The Coulomb field inside the nucleus will tend to push the protons towards the surface, an estimate of this effect (chapter 2.1 in \cite{Bohr69}) gives ground state isospin mixing probabilities that are expected to go like $Z^2 A^{2/3} / (T+1)$ for states of isospin $T$, and therefore to be largest at $N=Z$ and increase rapidly with the charge number. This pattern is seen experimentally \cite{Rama75} with one of the largest isospin impurities found in $^{64}$Ga, as discussed later.

At higher excitation energies the level distances decrease and
when states of different isospin, but the same spin and parity, are close in energy even small perturbations suffice to give appreciable mixing,
the $2^+$ doublet just below 17 MeV excitation energy in $^8$Be is a well-known example, but other cases are known such as the IAS in $^{31}$S \cite{Benn16}. A general overview of isospin mixing at higher excitation energies is given in \cite{Mitc10}.

\subsection{Excitation energy comparisons}
The difference in absolute binding energy can be removed to lowest order by making use of double-differences, i.e.\ by looking instead on excitation energies in two mirror nuclei. Asymmetries that still remain are known as the long established Thomas-Ehrman shift and the more recently explored Coulomb energy difference (CED) systematics.

Excitation energies of states when one system is close to the particle emission threshold can differ substantially (up to 700 keV) as first observed in light nuclei \cite{Thom52,Ehrm51}, an effect now called the Thomas-Ehrman shift. Put briefly: the low angular momentum states in the most proton-rich nucleus will extend further from the core and thereby have lower Coulomb energy than states of high angular momentum or deeper bound states. A more detailed explanation includes the occupation fraction of the single particle state, in the R-matrix description of the effect \cite{Bark91} the key quantities are the penetrability for the different orbits and the strength of their coupling to continuum channels.
The effect was first seen in $^{13}$N and $^{13}$C, but can of course be extended to higher isospin systems, one example is the significantly lower excitation energy for the excited $0^+$ state in $^{12}$O with respect to $^{12}$Be \cite{Suzu16}. Here also a systematic correlation was found between the scaled energy difference and binding energy, with stronger asymmetries seen for weakly bound states of low angular momentum.

The double energy-difference technique can also reveal effects in more bound systems, such as those explored in high-spin bands in intermediate mass nuclei. The Coulomb energy differences measured mainly in $T = 1/2$ and $T=1$ nuclei are described in the comprehensive reviews \cite{Bent07,Lenz09}.
The level schemes of $T_z = \pm 1/2$ nuclei are strikingly similar, but display systematic differences in the range of tens of keV to hundreds of keV.  Several contributing effects have been identified by now, as described in detail in \cite{Bent07,Zuke02}. They include Coulomb multipole effects (as nucleons recouple and thereby change relative distance),  the Coulomb monopole term when the nuclear radius changes e.g.\ along a rotational band, single-particle corrections such as the electromagnetic spin-orbit contribution, and finally a sizeable contribution from CSB strong interactions.

In extreme cases, the above effects can give rise to different energy ordering of the levels in two mirror systems, in particular in regions with high level density. This will, for example, give different decay possibilities.
In extreme cases even ground states may differ in mirror systems, one example being the $^{16}$F-$^{16}$N ground state inversion. Recently also $^{73}$Sr and $^{73}$Br were shown to have a ground state inversion \cite{Hoff20}, but as pointed out in \cite{Lenz20} this is in no way surprising in view of the systematics referred to above. 

\subsection{Transition rates}

Transition rates will in general depend both on the available energy for the transition and an ``intrinsic strength'' characterizing the degree of similarity between initial and final state. In pertubation theory, these parts correspond in Fermi's golden rule to the phase space factor and the matrix element squared. For strong decays one may, e.g.\ for very broad levels, have to employ reaction theories such as R-matrix theory to deal properly with a transition, whereas perturbation theory should work for electromagnetic and weak decays. However, there are examples in light nuclei where electromagnetic or weak decays lead to broad final states, and in such cases a combined description of the decay chain may be needed.

When clear experimental asymmetries are present in decay rates the question is whether the asymmetry is due to differences in the underlying interactions (e.g.\ the presence of second class currents in weak decays mentioned earlier) or whether they can be explained by structural differences induced e.g.\ by an underlying difference in binding energy.
As an extreme example, the two mirror radiative capture reactions $^7$Be(p,$\gamma$)$^8$B and $^7$Li(n,$\gamma$)$^8$Li are at a first glance very different at low energy: on top of the explicit Coulomb barrier in the former it is dominated by captures at extranuclear distances whereas the latter has significant internal contributions. Nevertheless, the theoretical descriptions of the two processes proceed along similar lines and there is no clear evidence that nuclear charge symmetry breaking is needed, see \cite{Bark06,Foss15} and references therein.

Weak decays are normally thought of as proceeding exclusively to well-defined states in the daughter nucleus, a notion that is obviously true for transitions to narrow final states, but may not be appropriate if broad final states and/or loosely bound initial states are involved. In extreme cases it has been suggested \cite{Riis15} that beta decays may proceed directly to the continuum, in a similar way as Coulomb break-up may proceed (compare with the inverse process, the direct radiative decay mentioned in the previous paragraph).

As pointed out in the introduction, an alternative use of the term ``mirror beta transition'' is to designate beta transitions between $T=1/2$ mirror nuclei. Such transitions have turned out to be useful in precision studies of weak interactions, see \cite{Navi22,Seve21} and references therein. We shall here only deal with situations where two mirror nuclei both decay weakly into final states, i.e.\ we only look at systems with $T \geq 1$.

\subsection{Matrix elements}
The partial halflife $t$ for a beta transition is related to the beta strengths $B(F)$ and $B(GT)$ (the reduced matrix elements squared) via the standard relation
\begin{equation}   \label{eq:ft-relation}
  ft = \frac{\mathcal{T}}{B(F) + \left(\frac{g_A}{g_V}\right)^2B(GT)} \;,\;\;
  \mathcal{T} = \ln 2 \frac{2 \pi^3\hbar (\hbar c)^6}{g_V^2 (m_e c^2)^5} 
\end{equation}
where $g_V$ and $g_A$ are the vector and axial-vector weak coupling constants with the experimental ratio $g_A/g_V = -1.2754(13)$ \cite{PDG20} and the experimental value for $\mathcal{T}$ is 6144(4) s from superallowed Fermi decays \cite{Hard20}.
(Note that different definitions of the Gamow-Teller beta strength occur in the literature, in some the factor $g_A/g_V$ is included in $B(GT)$.)
The ft-values for allowed decays range from around $10^3$ (super-allowed transitions, corresponding to a beta strength of several units) to around $10^7$ and are therefore often quoted on a logarithmic scale as log(ft). For the mirror comparisons several measures have been used, e.g.\
\begin{equation}
  \delta = ft_+/ft_- -1
\end{equation}
or
\begin{equation}
  \Delta' = \sqrt{B(GT)_-} - \sqrt{B(GT)_+}.
\end{equation}
The quantity that will be employed below 
\begin{equation}
  \Delta = \mathrm{log(ft)}_+-\mathrm{log(ft})_-
  = \log\left( \frac{\mathrm{ft}_+}{\mathrm{ft}_-} \right)
\end{equation}
is closely related to $\delta$. In all cases a positiv asymmetry parameter implies that the $\beta^-$ decay has a larger strength.

Should one focus the attention on transition asymmetries in mirror systems where the matrix elements of the transitions are large or small? The experimental precision will depend on which specific nucleus is being studied, but may give a preference for strong transitions. It then often comes down to what the precision of theoretical models will be. Some cases are very well understood, the obvious example is pure Fermi super-allowed beta transitions that to lowest order only depends on the isospin quantum numbers. Here even small asymmetries may give meaningful tests of the basic theory, as will be discussed in detail in section \ref{sec:beta}.

In the opposite limit some transitions are allowed by selection rules, but are nevertheless very weak. If this is due to more or less accidental cancellations, the CSB effects could be easier to extract. A well-known example is given by the $^{14}$C and $^{14}$O transitions to the $^{14}$N ground state. The discussion below of the current theoretical understanding of the large experimental asymmetry observed in this system will illustrate the potential as well as the challenges of this class of mirror transitions.
It should in this connection be noted that higher-order transitions (e.g. first-forbidden decay for beta transitions, E2 electromagnetic transitions etc.) in principle could also give good symmetry tests as long as the theoretical understanding of the matrix elements is sufficiently advanced. The case of the first forbidden-transitions for $A=17$ will be discussed later.

The precision of a theoretical calculation may often be estimated by considering more data for the system in question than just the mirror beta transitions. This could include \cite{Wilk69} excitation energies for more levels, other beta transitions and some gamma transitions.

\section{Beta decay}\label{sec:beta}

\begin{figure}
 \resizebox{0.5\textwidth}{!}{ \includegraphics{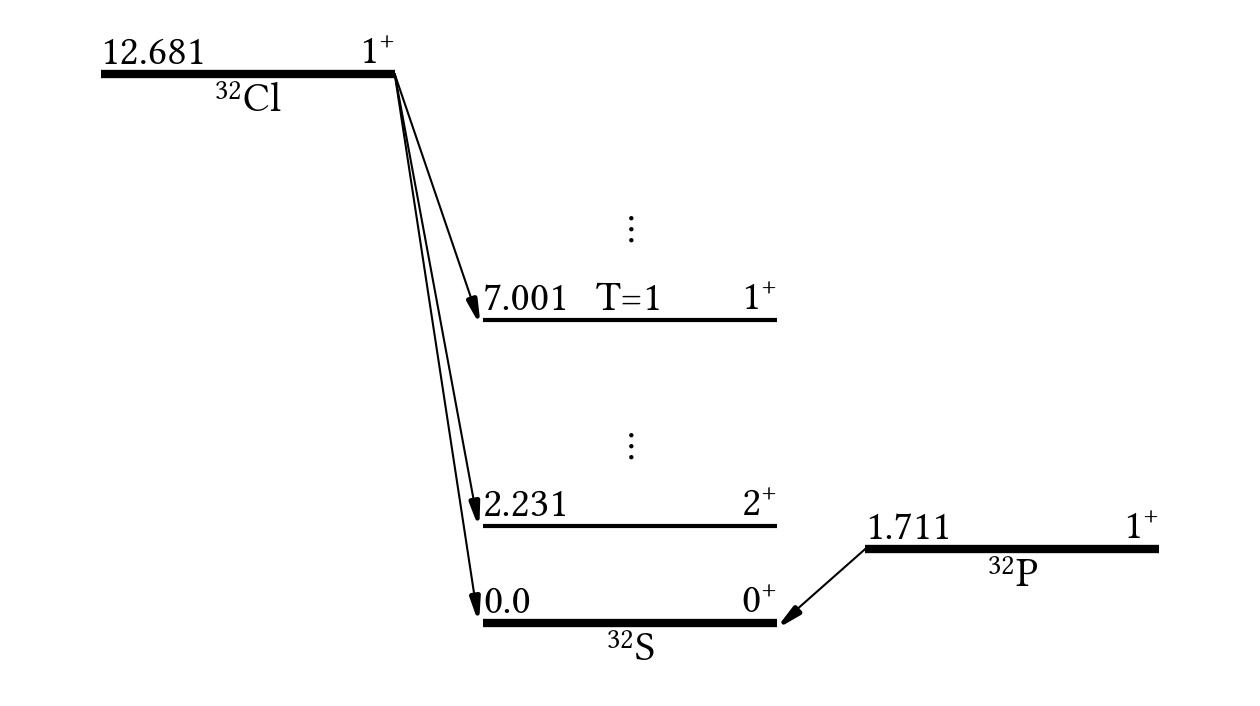} }
 \resizebox{0.5\textwidth}{!}{ \includegraphics{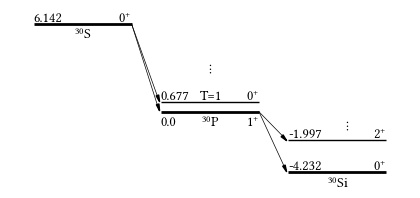} }
\caption{Schematic decay schemes for the two $T=1$ systems with mass number 32 (top) and 30 (bottom).}
\label{fig:A32-30}       
\end{figure}

Weak interactions are mainly probed in beta decays, with $\beta^-$ decays occuring for neutron rich nuclei and $\beta^+$ as well as EC (nuclear electron capture) for proton rich nuclei. Except for $^3$H (and the neutron), the mass
asymmetry makes $Q_{EC} > Q_{\beta^-}$ for a pair of mirror nuclei. The shifting of beta stability towards more neutron-rich nuclei implies that nuclei with $N=Z$ will no
longer be the most stable isobar for $A > 20$ (40) for $Z$ odd (even --- the difference is due to the odd-even effect). Figure \ref{fig:A32-30} illustrates this for the two triplets of nuclei with mass numbers 30 and 32, note how for the $N=Z$ nucleus the energy difference between the lowest $T=0$ and $T=1$ level differs drastically in the two cases. For $A=32$ both decays proceed towards the $N=Z$ nucleus (in this particular case the Q-value for $\beta^-$ decay is below the excitation energy of the first excited level), whereas for $A=30$ both decays proceed towards more neutron-rich nuclei. By using the principle of detailed balance, we can still deduce the log(ft) value for the decay of $^{30}$Si to $^{30}$P and therefore do a ``mirror decay comparison'', but since initial and final spins may differ we must correct by the factor $(2J_i+1)/(2J_f+1)$ (in the case shown, a factor of 3). Note that we in such cases only can investigate decays that proceed to the ground state of the $N=Z$ nucleus (and, in a few cases, to a long-lived isomeric state that has a measurable beta decay branch).
For $T=3/2$ a similar situation occurs above A=37, but as discussed below there are not yet sufficient data on the relevant decays in this region.

Fermi transitions proceed mainly within isospin multiplets and will therefore occur for $Z > N$ (here also part of the Gamow Teller Giant Resonance may become accessible for very proton-rich nuclei). Halflives are therefore short and will in fact be below one second for the proton-rich nucleus in a mirror system, except for some of the $T=1$ systems up to $^{30}$S.
Due to the short halflives we mainly have data on allowed mirror transitions, but first-forbidden transitions to low-lying excited states are also potentially visible. First-forbidden decays are only expected when the last protons and neutrons fill orbits with different parity, which happens for nuclei with $Z$ just above 8, 20 or 40. Above $Z=8$, $^{17}$Ne is the only such nucleus, above $Z=40$ there are currently no known cases, whereas there are a handful above $Z=20$, although none where there currently is sufficient data. Second forbidden transitions will proceed between states of the same parity, but are in practice not possible to see from the proton-rich nucleus. The data presented below  do not include such cases, that include e.g.\ $A=10$ where the $^{10}$Be decay is a $0^+ \rightarrow 3^+$ transition whereas only allowed transitions are observed in the $^{10}$C decay.

One can note that there are, apart from the super-allowed Fermi transitions shown below in Table \ref{tab:0to0}, only a few cases where mirror comparisons involve transitions with a large strength, say a log(ft) below 4. These cases appear mainly in light nuclei.

Could other weak interaction processes circumvent the Q-value limitation? Muon capture, where a $\mu^-$ is captured by a nucleus,
has a much higher $Q$-value than the corresponding electron capture, but proceeds in the same direction. Experiments on muon capture have so far had limited final state resolution \cite{Meas01},
the high available energy furthermore lessens the difference between allowed and forbidden
transitions. The positive muon $\mu^+$ is not captured in atoms, but may
induce a weak reaction on a target nucleus, similar to what
neutrinos and anti-neutrinos would. The cross sections are very small,
so measurements are not yet a practical possibility.
In the (far) future one could imagine $\mu^-$ capture on radioactive beams and $\mu^+$ reactions on stable nuclei as a way of accessing, for two mirror nuclei, the major part of the beta strength including the Gamow-Teller Giant Resonance (GTGR).
Well-chosen nuclear charge-changing reactions \cite{Fuji11} may give that information sooner, once they can be applied to radioactive beams as well.

\subsection{Example: Fermi transitions}
The superallowed Fermi transitions have a beta strength that to lowest order is simply $2T$. The pure Fermi transitions,  $0^+ \rightarrow 0^+$ decays in $T=1$ triplets, have for many decades been used to extract a value for the weak interaction strength in nuclei, see \cite{Navi22,Hard20} for recent overviews of this subject. (Recently the decays between mirror nuclei in $T=1/2$ doublets \cite{Navi22,Seve21} have also been put to use in this connection.) This group of decays are among the most carefully studied beta decays, experimentally as well as theoretically, and give an impressive accuracy and consistency. The quoted reviews give more details on the weak interaction aspects of these decays; the objective here is to look into cases where both $0^+ \rightarrow 0^+$ decays in an isospin triplet have been measured, the known (and some possible future) cases are listed in Table \ref{tab:0to0}. The most neutron-rich decays are better known experimentally and the extracted log(ft) values agree well even without theoretical corrections, although there is an increasing trend for higher masses. The proton-rich decays are more challenging experimentally and not yet on the same level of precision, they tend to have slightly larger (uncorrected) log(ft) values. The theoretical well-developed understanding of the decays can throw light on the beta asymmetry in this case. (Note that conversely the requirement of CVC symmetry after corrections has recently been employed in order to select a more consistent set of theoretical corrections \cite{Park14}.)

The interest here is in the nucleus-dependent and isospin-symmetry-breaking corrections. The contribtion from radiative corrections ($\delta_{NS}$ in \cite{Hard20}) is of order $10^{-3}$ and can be neglected here. The isospin-symmetry-breaking term ($\delta_C$) includes isospin/configuration mixing and incomplete radial overlap of wavefunctions, the latter the largest effect, and with a sum that range from 0.2\% to 1.7 \%, increasing with mass (and charge) number. It has traditionally been calculated in shell models employing Woods-Saxon potentials, but other theoretical frameworks have also been discussed, see the discussion in \cite{Hard20,Xaya22,Auer22}. Most of the correction is due to Coulomb effects, but \cite{Xaya22} finds that charge symmetry breaking nuclear terms increases the radial overlap term by between 10 \% and 30 \%. The isospin mixing term can in principle be tested experimentally by observing decays to other $0^+$ states, see e.g.\ the recent detailed measurement of the $^{62}$Ga decay\cite{MacL20}.

\begin{table}
\caption{The log(ft) values for $0^+ \rightarrow 0^+$ decays in $T=1$ triplets. Data are taken from the ENSDF and XUNDL databases at nndc.bnl.gov}
\label{tab:0to0}       
\begin{tabular}{lllll}
\hline\noalign{\smallskip}
$A$ & $\;\;\;\;T_z$ & \multicolumn{2}{c}{ log(ft)} & $\Delta$ \\
 & $-1,0,1$ & $-1\rightarrow 0$ & $1 \rightarrow 0$ $^{a}$ & \\
\noalign{\smallskip}\hline\noalign{\smallskip}
  26 & Si, Al$^m$, Mg & 3.4848(9) & 3.48281(6) & 0.0020(9) \\
  34 & Ar, Cl, S & 3.4846(12) & 3.48448(12) & 0.0001(12) \\
  38 & Ca, K$^m$, Ar & 3.487(1) & 3.48482(13) & 0.002(1) \\
  42 & Ti, Sc, Ca & 3.495(11) & 3.4845(2) & 0.010(11)\\
  46 & Cr, V, Ti & 3.492(9) & 3.4845(4) & 0.007(9) \\
  50 & Fe, Mn, Cr & 3.49(1) & 3.4846(2) & 0.01(1) \\
  54 & Ni, Co, Fe & 3.501(15) & 3.48460(17) & 0.016(15) \\
  62 & Ge, Ga, Zn & - & 3.48820(13) &  \\
  66 & Se, As, Ge & - & - & \\
  70 & Kr, Br, Se & - & - & \\
  74 & Sr, Rb, Kr & - & 3.4895(8) & \\
  78 & Zr, Y, Sr & - & - & \\
  82 & Mo, Nb, Zr & - & 3.59(8) & \\
  86 & Ru, Tc, Mo & - & 3.84(8) & \\
  90 & Pd, Rh, Ru & - & (3.6) & \\
  94 & Cd, Ag, Pd & - & - & \\
\noalign{\smallskip}\hline
\end{tabular}

$^{a}$Deduced from the inverse transition
\end{table}

Several cases with large isospin-symmetry breaking effects have also been found (but the mirror transition is not energetically allowed in these cases), namely the decay of $^{32}$Cl where the experimental correction was 5.4(9) \% \cite{Melc12} and the decay of $^{31}$Cl where large mixing of the IAS with a closely lying level was discovered \cite{Benn16}. These cases may provide information on the CSB interactions, as shown by \cite{Kane17} that also in the analysis included data for Coulomb energy differences.

A conceptually different way of addressing the nuclear mismatch corrections for Fermi decays was put forward within the R-matrix theory \cite{Bark92a,Bark94a}. This formalism may include continuum effects better than standard methods and could be an alternative starting point for the decays that are situated close to the proton drip line.
A very similar calculational procedure is of course also possible for Gamow-Teller decays \cite{Bark92b,Bark94b}.

\subsection{Example: $^{14}$C and $^{14}$O decays}
The strongly suppressed allowed decay of $^{14}$C into the $1^+$ ground state of $^{14}$N has been an often discussed challenge for nuclear theory for decades. An overview of much of the early literature can be found in \cite{Genz91,Robs11} that both employ a phenomenological model in L-S coupling where the lowest $T=1$ states turn out to be mainly $L=0$ whereas the $T=0$ $^{14}$N ground state has mainly $L=2$ (and 1) components.

For a full elucidation of the problem one needs to consider more experimental data, in particular the mirror decay 
of $^{14}$O to the ground state of $^{14}$N. Both decays are unusually slow for an allowed transition with log $ft$ values for $^{14}$C and $^{14}$O of  9.040(3) and 7.280(8) (7.365(13) from the latest measurement \cite{Voyt15}). It is noteworthy that the decay of $^{14}$O into the second $1^+$ state in $^{14}$N has a very low log(ft) of 3.131(17), the $^{14}$C(p,n) reaction shows that the same pattern exists for $^{14}$C \cite{Fuji20}.

With the small matrix elements of the ground state transitions higher order terms, e.g.\ weak magnetism, may also contribute and the shape of the beta spectrum will be modified (details can be found in \cite{Genz91,Town05} that also point out that renormalized operators must be used). This in turn can affect the value that is extracted for the matrix elements, different evaluations \cite{Robs11,Town05} yield a ratio between the matrix elements squared in the range 20 to 60. Out of the many theoretical studies of these decays, some have focussed on explaining why the matrix elements are so small, some have also attempted to explain the very large difference between the two decays. The requirement that the absolute rate should be explained for $^{14}$C implies that the rate for $^{14}$O should be calculable to a few percent precission, which is probably demanding too much from the theoretical models, but one can at least hope for an explanation for the order-of-magnitude of the decay rates.

The pattern of very high and very low log(ft) for the two $1^+$ states is in \cite{Fuji20} related to a simplified picture with effective two nucleon (or hole) transitions, and is shown to be similar to the decays of other closed shell plus two particles/holes, such as $^6$He, $^{18}$O and $^{42}$Ca. The difference between the mirror decays of $^{14}$O and $^{14}$C is related to the small difference in the p-orbit energy splitting in the two nuclei, both nuclei are situated close to a zero point for the beta strength plotted as a function of the p-orbit energy splitting.
A cluster model \cite{Robs11} was able to accomodate many of the experimental data on the system, predicted at zeroth order that the ground state decays are forbidden and attritubed the difference in decay rate in the two mirror nuclei to the (small) electromagnetic spin-orbit effect that naturally has opposite sign in the two cases.
Note that many recent papers have focussed on $^{14}$C and do not explain the difference to $^{14}$O (and most do not consider the shape-corrections). Several ab initio calculations have explored the ability of 3NF (or density dependence) and two-body currents to explain the cancellation \cite{Holt08,Mari11,Ekst14}.

\section{The experimental situation}
The currently available experimental data on beta asymmetries will be presented in tabular and graphical form after a few introductory general remarks. Some of the systems are then discussed in more detail, in particular the ones where large asymmetries seem to occur. This discussion is done separately for the following three mass regions: 
(i) Mass 8, 9 and 12 -- these nuclei have small $Z$ and high $Q$-values and therefore almost symmetric decays. However, the decays often involve broad or otherwise complicated final states. This region will be discussed last.
(ii) From mass 13 to around mass 40 -- here the final states are narrower and there are fewer conceptual complications. Most of the known mirror decays belong to this group.
(iii) Above mass 40 -- the receding proton dripline and the shift of beta stability away from $N=Z$ (cf.\ Fig.\ \ref{fig:A32-30}) gradually reduces the number of possible mirror transitions. Among those
that are possible there is currently mainly data on $0^+ \rightarrow 0^+$ mirror decays.

For odd-odd $N=Z$ nuclei the ground state will for $A$ above 34 often have $T=1$ and the two ground state to ground state transitions in the triplet are then both superallowed $0^+$ to $0^+$ transitions. Table \ref{tab:0to0} gives the current status of such possible mirror transitions including the two cases where the $T=1, T_z=0$ state is an isomer whose beta decay is also known. As mentioned earlier, the most proton-rich transition is generally less well known (for the heaviest nuclei not yet determined) than the decay of the $T_z = 0$ nucleus that are among those used  \cite{Hard20} to determine the value of $\mathcal{T}$ in eq. (\ref{eq:ft-relation}).

To extract reliable values for the beta strength, experimental data on Q-values, excitation energies, branching ratios and overall halflife is needed. As practitioners are aware, constructing an accurate decay scheme can be challenging. Many of the nuclei considered below will have sufficiently high Q-values to display prominent beta-delayed proton (or alpha particle) emission, so that these particles must be detected along with emitted gamma-rays. Competition between particle and gamma emission can occur, an important special case is for the Isobaric Analogue State (IAS) that will be fed prominently in nuclei that are not too proton rich (the IAS will for higher isospin multiplets be situated at higher excitation energy and therefore not dominate the decay as it can for $T=1/2$ or 1). Gamma-emission from the IAS can be a significant decay branch and must be considered \cite{Roec90}. An example is the decay of $^{17}$Ne discussed below, where gamma-branches from the IAS affected the deduced feeding to the first excited state.

The ground state feedings can be quite difficult to measure, but can be obtained via direct measurements of the beta spectrum; this is challenging when the branching is low, a recent example is the measurement \cite{Kirs19} of the ground-state transition in the $^{20}$F decay where the log(ft) was found to be 10.89(11). This affects in particular nuclei with large Q-windows and limits precision for many ground state transitions, thereby complicating mirror comparisons in heavier nuclei where the decay direction for the neutron-rich decay is inversed and only the ground state branch may be compared. If it happens that the ground state transition is forbidden, there will at least for the proton-rich nucleus often be several allowed transitions at higher excitation energy and measurements of the ground state feeding can be practically impossible. Such cases are excluded from the comparisons below.

\subsection{Data on mirror beta decays}
All data in the following tables and figures are taken from the Evaluated Nuclear Structure Data File database (ENSDF), occasionally supplemented with data from the Experimental Unevaluated Nuclear Data List (XUNDL), both available at the National Nuclear Data Center (online at nndc.bnl.gov, consulted October 2022). In selected cases the original papers will be discussed in later subsections. The presentation will proceed via increasing isospin of the decaying mirror nuclei.

\begin{figure}
  \resizebox{0.5\textwidth}{!}{ \includegraphics{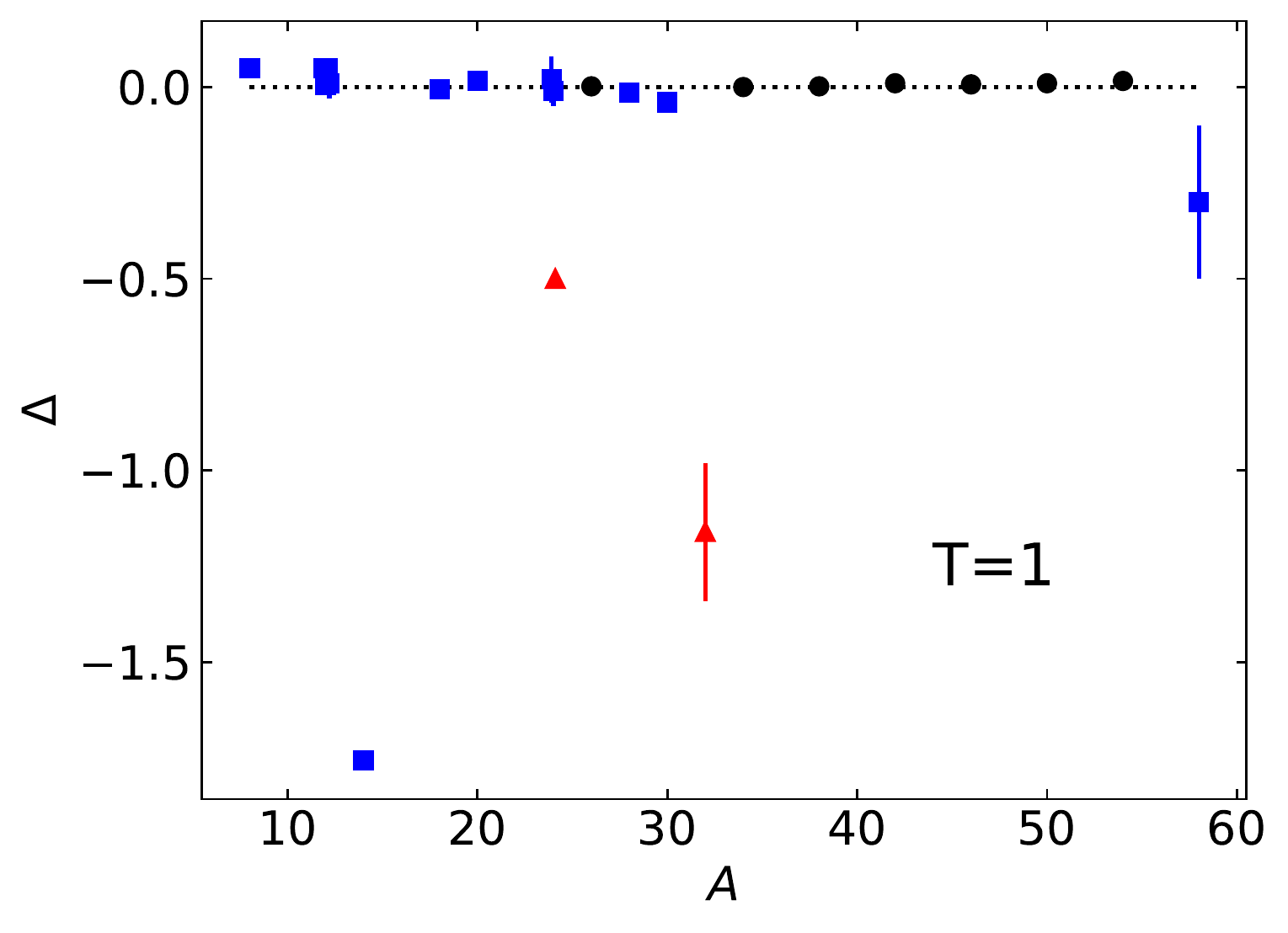} }
\caption{The difference $\Delta$ of log(ft) values is displayed versus mass number for $T=1$ mirror systems. Black circles denote $0^+ \rightarrow 0^+$ decays, blue squares other transitions. Decay asymmetries marked by a red triangle could be unreliable and are discussed in the text.}
\label{fig:T1}    
\end{figure}

The mirror decays with $T=1$ that do not belong to the group of $0^+ \rightarrow 0^+$ transitions (listed above in Table \ref{tab:0to0}) are collected in Table \ref{tab:T1}. All the $T=1$ transitions are displayed in Fig.\ \ref{fig:T1}. Here and in the following tables, when more mirror transitions are known for a given system they will be listed in order of increasing excitation energy of the final level.

\begin{table}
\caption{The log(ft) values for $T=1$ triplets except $0^+ \rightarrow 0^+$ decays. Data are taken from the ENSDF and XUNDL databases at nndc.bnl.gov}
\label{tab:T1}       
\begin{tabular}{lllll}
\hline\noalign{\smallskip}
$A$ & $\;\;\;\;T_z$ & \multicolumn{2}{c}{ log(ft)} & $\Delta$ \\
 & $-1,0,1$ & $-1\rightarrow 0$ & $1 \rightarrow 0$ & \\
\noalign{\smallskip}\hline\noalign{\smallskip}
  8 & B, Be, Li & 5.77 & 5.72 & 0.05(1) \\  
  12 & N, C, B & 4.1106(7) & 4.0617(5) & 0.0489(8) \\  
     & & 5.148(8)  & 5.143(7) & 0.005(11) \\  
     & & 4.622(10) & 4.572(17) & 0.05(2) \\ 
     & & 3.924(11) & 3.91(4) & 0.01(4) \\ 
  14 & O, N, C & 7.280(8) & 9.040(3) & $-$1.756(9) \\
  18 & Ne, F, O & 3.0885(11) & 3.0944(15) $^{a}$ & $-$0.006(2) \\
  20 & Na, Ne, F & 4.987(7) & 4.9697(11) & 0.017(7) \\  
  24 & Al, Mg, Na & 6.13(6) & 6.11(1) & 0.02(6) \\
    & & 6.59(4) & 6.60(2) & $-$0.01(4)\\
  24 & Al$^b$, Mg, Na$^b$ & 5.30(2) & 5.80 & $-$0.5 \\  
  28 & P, Si, Al & 4.851(5) & 4.8664(4) & $-$0.015(5) \\
  30 & S, P, Si & 4.322(11) & 4.3622(8) $^{a}$ & $-$0.040(11) \\
  32 & Cl, S, P & 6.74(18) & 7.9002(2) & $-$1.16(18) \\
  58 & Zn, Cu, Ni & 4.1(2) & 4.393(3) $^{a}$ & $-$0.3(2) \\
\noalign{\smallskip}\hline
\end{tabular}

$^{a}$Deduced from the inverse transition

$^{b}$The $1^+$ isomer
\end{table}

\begin{figure}
  \resizebox{0.5\textwidth}{!}{ \includegraphics{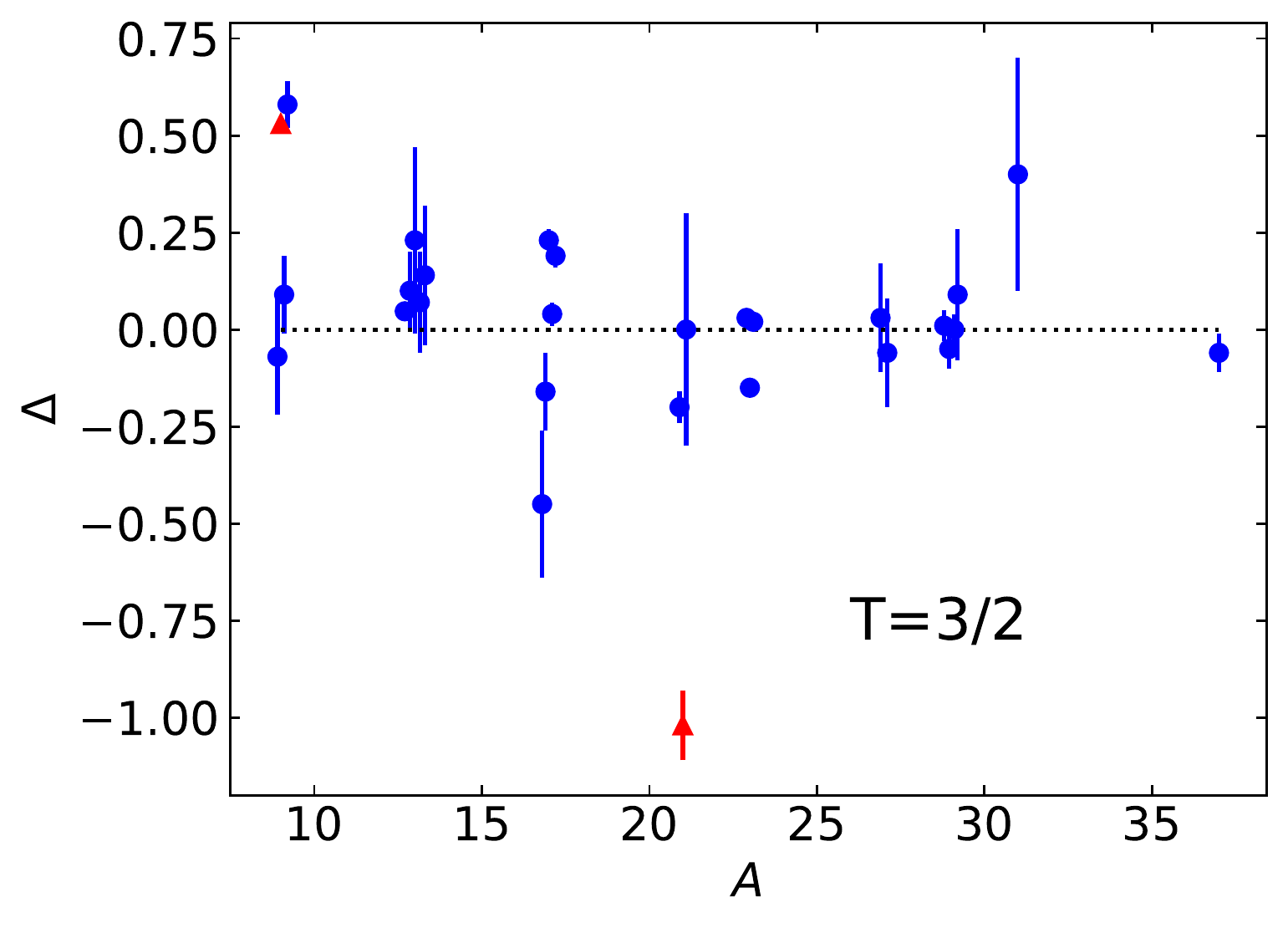} }
\caption{The difference $\Delta$ of log(ft) values is displayed versus mass number for $T=3/2$ mirror systems.  Decay asymmetries marked by a red triangle could be unreliable and are discussed in the text.}
\label{fig:T32}    
\end{figure}

The mirror decays with $T=3/2$ are collected in Table \ref{tab:T32} and displayed in Fig. \ref{fig:T32}. In some case it has been so difficult to measure low-lying transitions, in partuclar the ground state transition, that mirror symmetry was assumed in order to estimate transition intensity. Such cases are marked in the table and no asymmetry can be deduced at the moment.

\begin{figure}
  \resizebox{0.5\textwidth}{!}{ \includegraphics{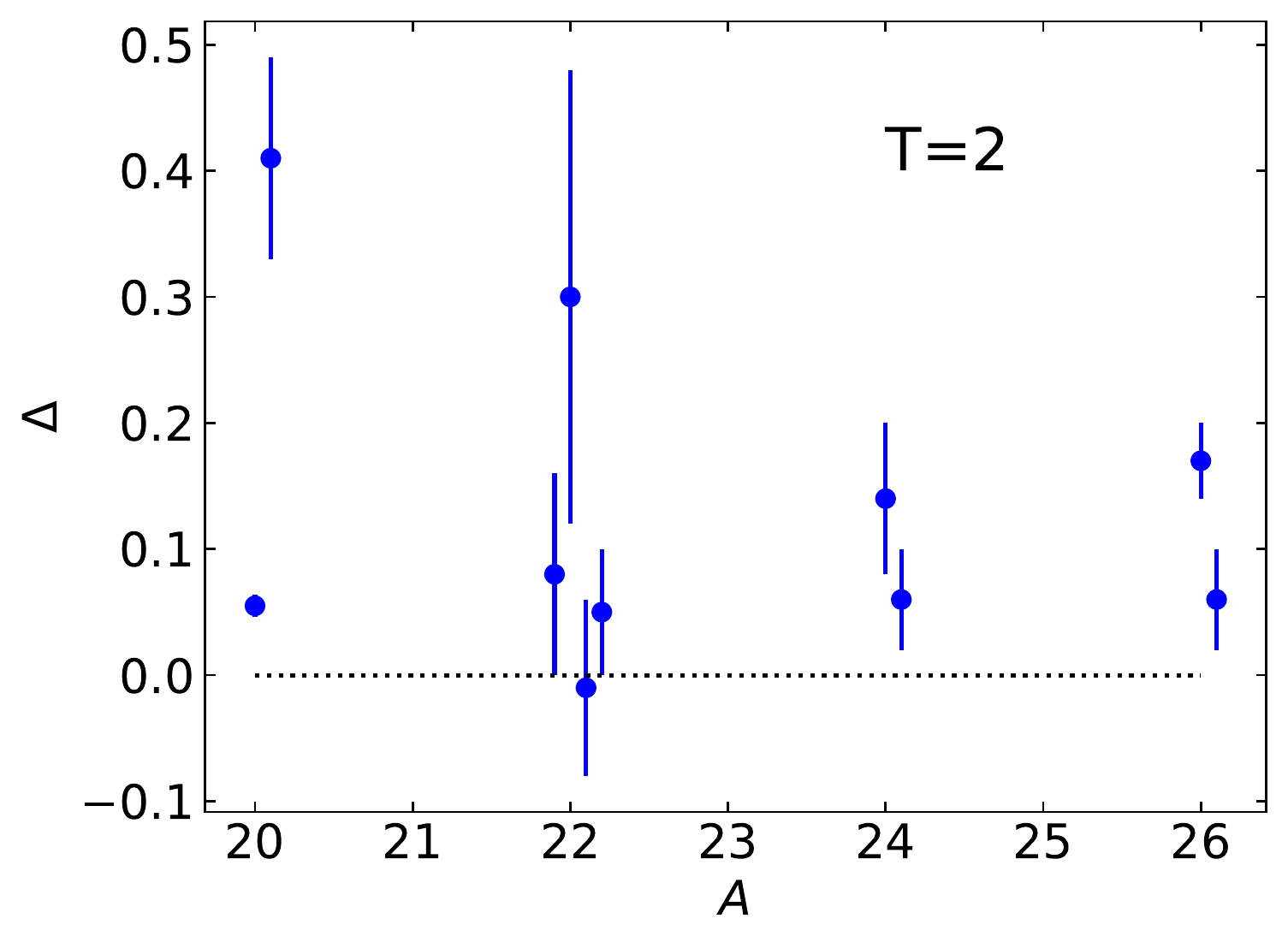} }
\caption{The difference $\Delta$ of log(ft) values is displayed versus mass number for $T=2$ mirror systems. The data for mass 26 are the two lowest transitions  as reported in \protect\cite{Pere16}.}
\label{fig:T2}    
\end{figure}

The mirror decays with $T=2$ are listed in Table \ref{tab:T2} and transitions are displayed in Fig. \ref{fig:T2}. As could be expected there are significantly fewer transitions measured compared to the previous groups. 

Finally, the mirror decays with $T=5/2$ only occur for cases where the $T_z = -5/2$ nucleus has an even $Z$. For odd $Z$, nuclei with $T_z= -5/2$ are unbound to particle emission. 
Also for $T=3$ only a few multiplets are particle bound among which only one case, with $A=22$, has experimentally established mirror decays. In both cases, the data will be discussed in the next subsections.

\subsection{Mass numbers 13 to 40}

\emph{T=1.} For $A=22,36,40$ the spin differences in the ground state decays are too high to allow mirror comparisons to be performed. The case of $A=14$ has the largest recorded decay asymmetry and was discussed in detail above.

The asymmetry for the decays of the $1^+$ isomers in $^{24}$Al and $^{24}$Na to the low-lying states in $^{24}$Mg could be due to incomplete information on the decays. The data on the $^{24}$Na isomer decay stems from its discovery experiment \cite{Drop56} and are quite uncertain, the only other observation seems to be from an experiment where the magnetic moment was measured \cite{Heit80}. The uncertainty on the quoted log(ft) value must be at least 0.3. The two latest experiments on the $^{24}$Al isomer decay \cite{Honk79,Nish11} give quite different intensities for the ground state branch, so new experiments are probably needed for both isomers to obtain a reliable asymmetry.

The apparent large asymmetry for $A=32$ depends on the $^{32}$Cl decay to the $^{32}$S ground state that in ENSDF has a quoted log(ft) of 6.74(18) and an intensity of the branch of 1.0(4) \%. The original paper \cite{Armi68} gave an intensity of $1.0^{+0.2}_{-0.5}$ \%, noted the mirror asymmetry and also commented on the similarity to the situation for $^{14}$C as well as that the beta spectrum here also may be nonallowed. However, the quoted intensity is only two standard deviations from zero, so the asymmetry is therefore not really established experimentally.

\begin{table}
\caption{The log(ft) values for $T=3/2$ quadruples. Data are taken from the ENSDF and XUNDL databases at nndc.bnl.gov}
\label{tab:T32}       
\begin{tabular}{lllll}
\hline\noalign{\smallskip}
$A$ & $\;\;\;\;\;T_z$ & \multicolumn{2}{c}{ log(ft)} & $\Delta$ \\
 & $-\frac{3}{2},-\frac{1}{2},\frac{1}{2},\frac{3}{2}$ & $-\frac{3}{2}\rightarrow -\frac{1}{2}$ & $\frac{3}{2} \rightarrow \frac{1}{2}$ & \\
\noalign{\smallskip}\hline\noalign{\smallskip}
  9 & C, B, Be, Li & 5.318(13) & 5.325(8) & -0.07(15) \\ 
     & & 5.22(9) & 5.13(5) & 0.09(10) \\ 
     & & 5.87(6) & 5.34(9) & 0.53(11) \\ 
     & & 3.14(5) & 2.56(4) &  0.58(6) \\ 
  13 & O, N, C, B & 4.081(11) & 4.034(6) & 0.047(13) \\ 
    & & 4.55(9) &  4.45(5) & 0.10(10) \\ 
    & & 5.56(22) & 5.33(10) & 0.23(24) \\ 
    & & 4.66(10) & 4.59(9) & 0.07(13) \\ 
    & & 5.09(11) & 4.95(14) & 0.14(18) \\ 
  17 & Ne, F, O, N & 6.39(5) & 6.84(9) & $-$0.45(19)\\ 
    & & 6.94(4) & 7.10(9) & $-$0.16(10)\\ 
    & & 4.65(3) & 4.416(15) & 0.23(3) \\ 
    & & 3.895(24) & 3.851(13) & 0.04(3) \\ 
    & & 4.569(24) & 4.380(23) & 0.19(3) \\ 
  21 & Mg, Na, Ne, F & 6.09(8) & 7.11(5) & $-$1.02(9) \\  
    & & 4.82(3) & 5.02(3) & $-$0.20(4) \\  
    & & 4.48(3) & 4.5(3) & 0.0(3) \\ 
  23 & Al, Mg, Na, Ne & 5.30(2) & 5.27(1) & 0.03(2) \\
    & & 5.36(1) & 5.38(2) & $-$0.02(2) \\
    & & 5.67(1) & 5.82(2) & $-$0.15(2) \\
  25 & Si, Al, Mg, Na & 5.24(14) & 5.26 & \\
    & & 5.05(7) & 5.04 & \\ 
  27 & P, Si, Al, Mg & 4.76(14) & 4.7297(10) & 0.03(14) \\
    & & 4.87(14) & 4.9340(16) & $-$0.06(14) \\
  29 & S, P, Si, Al & 5.06(4) & 5.050(5) & 0.01(4) \\
    & & 5.73(4) & 5.733(13) & 0.00(4) \\
    & & 4.98(5) & 5.026(15) & $-$0.05(5) \\
    & & 6.20(15) & 6.11(7) & 0.09(17) \\
  31 & Cl, S, P, Si & (5.6) $^{a}$ & 5.5250(6) & \\
       & & 6.1(3) & 5.747(6) & 0.4(3) \\
  33 & Ar, Cl, S, P & 5.022(11) $^{a}$ & 5.022(7) & \\
  35 & K, Ar, Cl, S & 5.07(19) $^{a}$ & 5.0088(7) & \\
  37 & Ca, K, Ar, Cl & 5.04(5) & 5.1006(4) $^{b}$& -0.06(5)\\
\noalign{\smallskip}\hline
\end{tabular}

$^{a}$Estimated using the mirror transition

$^{b}$Deduced from the inverse transition
\end{table}

\emph{T=3/2.} 
Several transitions differ in strength between the mirror nuclei $^{17}$Ne and $^{17}$N. The main recent activity has focussed on the first-forbidden transition to the first excited $1/2^+$ states that experimentally \cite{Borg93,Ozaw98} are about twice as strong in $^{17}$Ne (the first entry for mass 17 in Table \ref{tab:T32}). Several calculations have explored to what extent this asymmetry is related to the halo nature of $^{17}$F $1/2^+$ state (or the suggested halo in $^{17}$Ne) \cite{Borg93} or to different occupations of orbits in the decaying nuclei \cite{Mill97} and the current conclusion \cite{Mich02} is that first-forbidden decays are too complex to be interpreted in a simple way. However, it is noteworthy that several higher-lying allowed transitions also are asymmetric, the main exception being the most intense transition to the $3/2^-$ level just below 5.5 MeV. A combined look at several decay branches may yield new information.

The large asymmetries given for $A=21$ may not be reliable. They come from the $^{21}$Mg decay to the $5/2^+$ levels at 3.54 MeV and 4.29 MeV and the quoted strengths come from an early experiment \cite{Sext73} that appears to not resolve all transitions, a more recent experiment \cite{Lund15} give instead $\Delta$ values for the three transitions in Table \ref{tab:T32} of $-0.66(17)$, 0.12(16) and 0.2(3). Newer data on this decay are likely to appear soon \cite{Jens22} and may clarify the situation.

\emph{T=2.} 
There are fewer cases known -- the heaviest odd-odd $T_z=-2$ nuclei are unbound, so $^{26}$P is the last such case -- and experimental uncertainties are in general larger. Among the cases where it will be impossible to measure both transitions are $A=36$, here the neutron-rich transition proceeds in the inverse direction $T_z= 1 \rightarrow 2$ and the ground state transition is second forbidden with log(ft) = 13.58(3). 

For mass 20, the two quoted transitions are to the two lowest $1^+$ levels in the daughter nuclei at around 1.0 and 3 MeV. Concerning the large asymmetry to the second level, there are four consistent experiments \cite{Piec95,Wall12,Lund16,Sun17} on the $^{20}$Mg decay performed under different conditions and one careful experiment \cite{Albu87} on $^{20}$O so there are no obvious problems with the data. Still, a confirmation of the $^{20}$O result would be valuable.

For mass 26, there are more than 20 transitions known both for the $^{26}$P decay \cite{Benn13,Pere16} and for the mirror $^{26}$Na decay \cite{Grin05}. Table \ref{tab:T2} only mentions the lowest ones and use values quoted in XUNDL for $^{26}$P, Table VI in \cite{Pere16} has a more detailed comparison of the lowest ten levels and notes several asymmetries $\delta$ significantly different from zero. From the log(ft) values in the paper the lowest $2^+$ state would have a $\Delta = 0.17(3)$. The asymmetry is in \cite{Pere16} tentatively attributed to $^{26}$P having a proton halo, a suggestion supported by shell model calculations.

For mass 28 there is a lack of data for the $^{28}$S decay to low-lying states in $^{28}$P.

\begin{table}
\caption{The log(ft) values for $T=2$ quintuples. Data are taken from the ENSDF and XUNDL databases at nndc.bnl.gov}
\label{tab:T2}       
\begin{tabular}{lllll}
\hline\noalign{\smallskip}
$A$ & $\;\;\;\;T_z$ & \multicolumn{2}{c}{ log(ft)} & $\Delta$ \\
 & $-2,-1,1,2$ & $-2\rightarrow -1$ & $2 \rightarrow 1$ & \\
\noalign{\smallskip}\hline\noalign{\smallskip}
  20 & Mg, Na, F, O & 3.789(9) & 3.7340(6) & 0.055(9)\\ 
    &   & 4.05(6) & 3.64(6) & 0.41(8) \\ 
  22 & Al, Mg, Ne, F & 4.87(8) & 4.79(1) & 0.08(8) \\ 
    & & 5.56(18) & 5.26(2) & 0.30(18) \\ 
    & & 5.33(7) & 5.34(2) & $-$0.01(7)\\ 
    & & 4.75(5) & 4.70(2) & 0.05(5) \\ 
  24 & Si, Al, Na, Ne & 4.49 (6) & 4.35(1) & 0.14(6) \\
    &   & 4.45(3) & 4.39(2) & 0.06(4) \\
  26 & P, Si, Mg, Na $^{b}$ & 4.9(1) & 4.71(1) & 0.2(1)\\  
     & &  $>$6.6 & 7.6(4) & \\  
     & &  5.8(1) & 5.87(1) & 0.0(1) \\  
  32 & Ar, Cl, P, Si & $>5.1$ & 8.21(6) & \\  
  46 & Mn, Cr, Ti, Sc & - & 6.200(3) & \\
  64 & Se, As, Ga, Zn & - & 6.65(2)  $^{a}$ & \\
\noalign{\smallskip}\hline
\end{tabular}

$^{a}$Deduced from the inverse transition

$^{b}$See the text
\end{table}

\emph{T=5/2.} Here only the $T_Z= -5/2$ nuclei with even $Z$ exists and beta decays. The experimental situation is as follows: the $T_z = 5/2 \rightarrow 3/2$ decays are generally established, except that the normalization of the $^{31}$Al decay is uncertain. The $T_z = -5/2 \rightarrow -3/2$ decays are all studied, but transitions to low-lying states are difficult to get on an absolute scale, although the ones for $^{27}$S are roughly consistent with the mirror decay \cite{Jani17}.

\emph{T=3.} The only case in this group is the decay of $^{22}$Si where transitions to the two lowest $1^+$ states in $^{22}$Al \cite{Lee20} have log(ft) values of 5.09(14) and 3.83(13), and the mirror decays of $^{22}$O give log(ft) values of 4.6(1) and 3.8(1) \cite{Weis05,Hube89} (the error bars are from ENSDF/XUNDF). The deduced values of $\Delta$ for the two transitions are 0.49(17) and 0.03(16). The large asymmetry is in \cite{Lee20} attributed to a proton halo structure in the first $^{22}$Al $1^+$ level due to low binding and an enhanced occupation of the proton s-orbit. It would be interesting to gather more information on this level.

\subsection{Mass numbers above 40}

For \emph{T=1} nuclei only a few cases are not included already in table \ref{tab:0to0}. For $A=58$ the $N=Z$ nucleus $^{58}$Cu has a $T=0$ $1^+$ ground state (breaking the systematics of table \ref{tab:0to0}); it is fed from $^{58}$Zn with a log(ft) of 4.1(2) and decays to the ground state of $^{58}$Ni with a log(ft) of 4.870(3), a difference mainly due to the spin weighting factor $2J+1$.
For $A=64$ the ground states of $^{64}$Se and $^{64}$Zn ($T=2$), $^{64}$As and $^{64}$Ga ($T=1$), and $^{64}$Ge ($T=0$) all have spin-parity $0^+$, so beta transitions between them would all be isospin forbidden Fermi transitions; nevertheless the decay of Ga into Zn is reported with a log(ft) of 6.65(2). This represents one of the largest known isospin mixings \cite{Rama75} among non-degenerate levels. It would clearly be of interest to measure the amount of isospin mixing for more ground-state transitions.

For \emph{T=3/2} nuclei only the ground state decays of the $T_z = -3/2 \rightarrow -1/2$ and $T_z = 1/2 \rightarrow 3/2$ transitions can be compared. Here there is not yet sufficient experimental knowledge: For mass numbers $A= 43-53, 61$ only the latter transition has been measured, for $A=65$ not even that, for $A=41, 55, 57$ the ground state transition is not allowed and in the remainder of the cases not all members are particle bound.

For \emph{T=2} there is one well-determined transition in $^{46}$Sc$\rightarrow$$^{46}$Ti, but the mirror transition in $^{46}$Mn$\rightarrow$$^{46}$Cr has not yet been observed. 

For \emph{T=5/2} nuclei the only cases where the $T_z = -5/2$ nucleus is particle bound are at $A= 43, 47, 51, 55$ and there is also here a lack of data for their decays. 

There are only a few proton rich nuclei with isospin \emph{T=3} and higher. For many of them the mirror decay cannot be measured or is forbidden, but the \emph{T=7/2} mirror pairs
$^{45}$Fe - $^{45}$K and $^{49}$Ni - $^{49}$Sc may give relevant data in the future once accurate measurements of the proton rich nucleus are doable.

\subsection{Mass numbers 8, 9 and 12}
This group of nuclei have for a long time been known to have asymmetric mirror decays and have been of interest in connection with second class current searches \cite{Wilk70,Wilk00}. A common feature of the three systems is that final continuum states including alpha particles play a significant role.

The \emph{mass 8} decays, of $^8$B and $^8$Li, proceed fully into unbound excited states of $^8$Be, so far only $2^+$ final states have been shown to contribute. The decay of both nuclei have been studied carefully in many experiments, partly in connection with weak interaction studies \cite{Vos15}, partly due to the role $^8$B plays in the solar neutrino spectrum, see \cite{Kirs11,Roge12} for details. The main decay goes to the broad 3 MeV level in $^8$Be, but more components are present at high energy, and different interpretations have been put forward, see \cite{Riis15} and references therein. The strength is known in both decays from below 1 MeV to close to 16 MeV excitation energy, which allowed Wilkinson and Alburger \cite{Wilk71} to extract the asymmetry parameter as a function of energy. If arising from second class current the asymmety would depend on transition energy, but it was found to be constant as function of energy with a magnitude $\delta = 0.107(11)$ (corresponding to $\Delta = 0.044(4)$). The asymmetry parameter can alternatively be extracted from fits to the spectra in line with what is done for other decays; the thorough analysis in \cite{Bark89} gives log(ft) values that vary by about 0.13 when the channel radius in the R-matrix fits is changed, but the difference in the two mirror decays is consistent with a $\Delta$ of 0.05(1). This is the value quoted in Table \ref{tab:T1}.

It has been suggested \cite{Riis14} that the beta strength for this and similar cases where the final spectra are continuous so that levels may interfere, complicating the extraction of individual log(ft) values, should be derived instead as a strength function by employing Eq.\ (\ref{eq:ft-relation}) in each energy bin. It may be worthwhile to repeat the Wilkinson-Alburger comparison for strength functions extracted from the recent high-quality data.

The \emph{mass 9} decays, of $^9$C into $^9$B and $^9$Li into $^9$Be, present similar problems, as all $^9$C decays (and half of $^9$Li decays) go to final three-body continuum states. It is noteworthy that the decays to the lowest levels in the daughter nuclei are close to being symmetric, whereas the decays to the broad $1/2^-$ level at 3 MeV and the strongly fed level at 12 MeV are clearly asymmetric, see Table \ref{tab:T32}. Several experiments have over the past decades tried to clarify this issue, but most suffer from limitations in the way the three-body final states are described. The latest experiments effectively detect the three particles for each decay, see \cite{Prez05} and references therein, but the $1/2^-$ level is difficult to separate from other levels so assumptions on the decay mechanism have been made to obtain the current strength values. Changing the point of view to consider a beta strength function (similar to the Wilkinson-Alburger analysis) requires care, as the level positions in the two final nuclei differ somewhat. The asymmetry for the $1/2^-$ level has been marked as potentially unreliable in Fig.\ \ref{fig:T32}, the asymmetry for the 12 MeV level has at the moment no theoretical explanation.

The \emph{mass 12} decays of $^{12}$N and $^{12}$B into $^{12}$C populate four narrow levels -- the ground state, the $2^+$ bound state, the Hoyle state at 7.65 MeV and the narrow 12.7 MeV level, all listed in Table \ref{tab:T1} -- and several broad distributions in the triple-alpha particle continuum, where there is pronounced interference between levels and the assignment of strength to each levels again is quite difficult. Several recent experiments have attempted to unravel the decay pattern, the latest published results appeared a decade ago \cite{Hyld09a,Hyld09b}, but newer data have been collected and should appear soon \cite{Refs16,Gad21}. Since this is a $T=1$ system as for mass 8, one can again extract the decay asymmetry as a function of energy in the continuum (see Fig.\ 2 in \cite{Hyld09a}), the asymmetry is within uncertainties constant between the Hoyle state and the 12.7 MeV level with a $\Delta$ parameter of order 0.07.

It is noteworthy that both mass 8 and 12 have a constant asymmetry as function of excitation energy (that also is consistent with the asymmetry for the two $0^+$ level for mass 12), this may indicate that the difference is due to the mother nuclei. If this is related to binding energy effects, it is striking that both $^8$B and $^{12}$N have low proton separation energy, the former is often considered a proton halo. Calculations \cite{Bark94b,Town73} reproduce the order-of-magnitude of the asymmetry for all three masses.

\begin{figure}
  \resizebox{0.5\textwidth}{!}{ \includegraphics{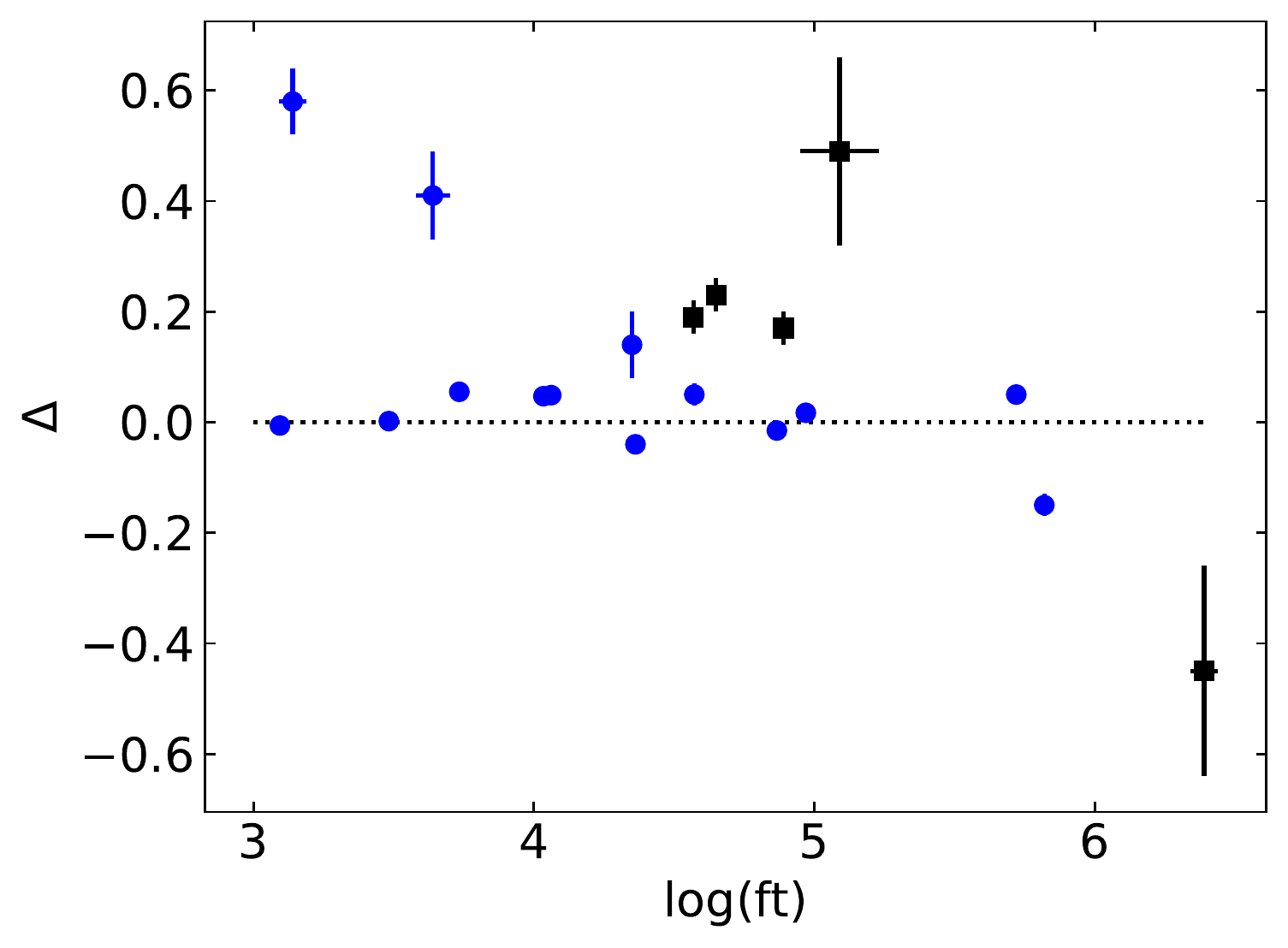} }
  \resizebox{0.5\textwidth}{!}{ \includegraphics{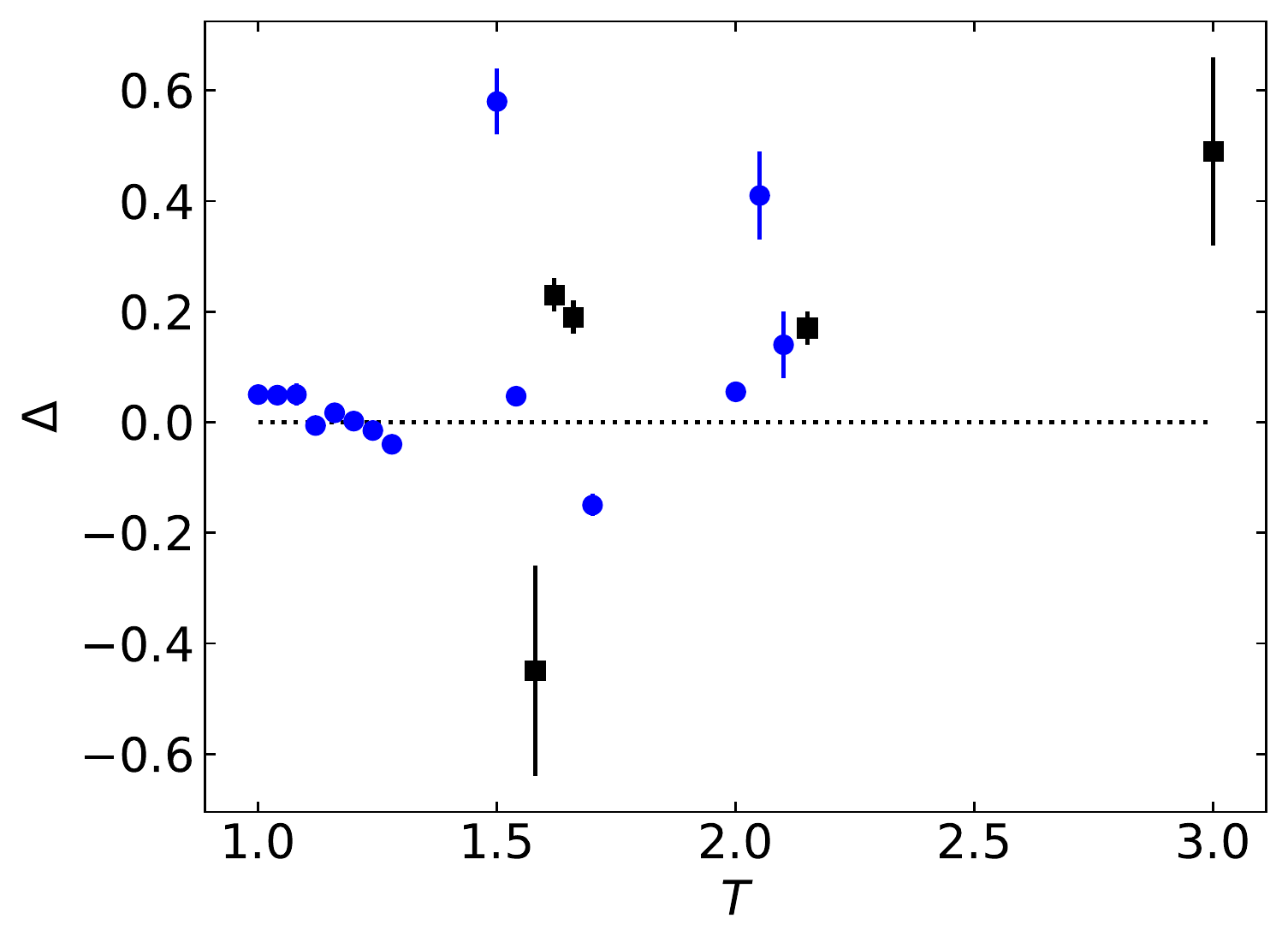} }
\caption{The difference $\Delta$ of log(ft) values is displayed versus (top) log(ft) and (bottom) $T$-value, in each group ordered according to increasing mass number.  Only cases that deviate from 0 by more than two standard deviations are included.  Decay asymmetries that could be unreliable and the $A=14$ system are not included. The black squares mark asymmetries for $^{17}$Ne, $^{22}$Si and $^{26}$P, possibly related to proton halos.}
\label{fig:asym}    
\end{figure}

\section{Discussion}

The first overall observation is that the experimentally observed beta asymmetries typically are small, in particular for accurately measured cases, and that there is a tendency for the asymmetries to scatter more the higher the isospin (i.e.\ the proton-neutron asymmetry) is, but since uncertainties naturally are higher for larger isospin, the larger scatter may simply be an experimental bias. 
There are experimental concerns for several of the reported large asymmetries, but also a fair number of asymmetries that are more than two standard deviations away from zero; these are collected in Fig. \ref{fig:asym}. Note here that asymmetries not only are better determined, but also are clearly smaller for $T=1$; this may be connected to the fact that both decays end up in the same daughter nucleus, whereas for higher isospin values the mirror decays go to different nuclei.
For all asymmetries $|\Delta| > 0.1$ the most proton-rich nuclei have isospin $T \geq 3/2$ and are (except for $^{23}$Al) situated at the proton dripline. As mentioned above, in the case of $^{17}$Ne, $^{22}$Si  and $^{26}$P proton halos have been (partially) invoked in order to explain the magnitude of the asymmetry. Here binding energy effects are expected to play a particularly important role. It is interesting that several of the precisely measured asymmetries cluster around 0.05, in particular for light nuclei. A better accuracy for more transitions with $T \geq 3/2$ is needed in order to identify systematic trends.

The number of mirror beta decays is naturally limited by the Coulomb effect that shifts the line of beta-stability as well as the proton dripline. A number of cases included here circumvents the shifting of the beta-stability line away from $N=Z$ by relating the ``not occuring $\beta^-$ decay'' to the inverse EC decay via the principle of detailed balance. However, only ground state transitions may be included in this way, and it is often experimentally challenging to measure the ground state branch of the corresponding EC decay with high accuracy. It would nevertheless still be interesting to have a few well measured cases of this type, as they would occur at higher mass values and therefore also higher $Z$ and so may provide tests for the physics causes of asymmetries. The currently well measured cases are some of the $T=1$ $0^+ \rightarrow 0^+$ cases and the $A=18$ triplet.

Having more mirror decays measured to an accuracy better than 1\% would be beneficial. Some specific cases where better data would help include:
\begin{itemize}
  \item $0^+ \rightarrow 0^+$ decays of the $T_z= -1$ nuclei in table \ref{tab:0to0}, it will be difficult but interesting to have better data for the heaviest cases where Coulomb corrections are largest  
  \item data on the differential asymmetry as a function of excitation energy for the $^9$Li and $^9$C decays
  \item the decay of the $^{24}$Na isomer
  \item completing the detailed investigation of the $^{32}$Cl decay \cite{Melc12} with a new measurement of the ground state branch
  \item more isospin mixing (forbidden Fermy decay) measurements for the $A = 64$ nuclei, most realistically $^{64}$Ge to $^{64}$Ga
  \item accurate data for one (or more) $T = 5/2$ mirror transitions
  \item detailed measurements on some of the nuclei above $^{40}$Ca
\end{itemize}
This wish-list constitutes a quite ambitious program.

\section{Conclusion}
\label{sec:final}

Mirror beta transitions will in the ideal case have identical beta strength, but several effects can contribute to an asymmetry. The second class currents in the weak interactions have been shown to give a negligible contribution. The strong interaction charge-symmetry breaking will give a noticable effect, but is dominated by binding energy effects (and other indirect effects of the electromagnetic interaction). A very similar conclusion was reached by Towner already half a century ago \cite{Town73}, namely ``that the binding energy effects are the most important, and yet at the same time are the most difficult to compute reliably''.
The theoretical situation has improved, in that modern shell model calculations now for two decades have been able to reproduce the Okamoto-Nolen-Schiffer anomaly as well as Coulomb energy differences, and also can reproduce some of the recently found large asymmetries \cite{Pere16,Lee20} as well as the correction terms to the $0^+ \rightarrow 0^+$ transitions \cite{Hard20,Xaya22}. However, they are unlikely to be able to explain the $^{14}$O-$^{14}$C asymmetry in detail, and there seems to be systematic problems with treating levels very close to or in the continuum. Frameworks to deal with such cases exist, e.g. the shell model embedded in the continuum employed for $^{17}$Ne \cite{Mich02}, but they have not been used systematically to describe the EC decays of proton-rich nuclei and the theoretical treatment of cases like $A=9,12$ where the final continuum states involves several particles remains an unsolved problem.

A large beta decay asymmetry indicates there are substantial structural differences in the two sets of mirror nuclei (most likely due to binding energy effects), but does not by itself inform about whether the diffrerence is in the mother or daughter nucleus, nor in the proton-rich or neutron-rich pair. It should always be interpreted in combination with other experimental information on the system.

Mirror Fermi and Gamow-Teller transitions differ greatly. For Fermi decays, the almost conserved isospin quantum number gives a model independent beta strength with only small corrections that then becomes the focus of investigations. One special concern is the amount of isospin mixing - this has now been handled experimentally in several cases, but it would be at least conceptually satisfying to see mirror isospin forbidden decays (it should be possible for $A=64$, but clearly not in the near future). The presently used procedure \cite{Park14} of selecting/fitting the theory to the assumption that CVC holds is of course optimal for weak interaction studies, but unsound in the present context and should not be employed when the focus is on beta asymmetries.
So far only small asymmetries have been observed, but heavier systems are likely to show larger effects. It could also be interesting to investigate more cases (not necessarily mirror decays) with larger corrections $\delta_C$. 

For Gamow-Teller decays the situation is more complex. There are no lowest order model independencies, the systems are more diverse and the asymmetries are here clearly larger, also with several cases being known for decades. As decays of more and heavier proton-rich nuclei have been explored more examples of asymmetries have turned up, the most striking is arguably the proton halo systems with their sizable binding energy effects.

Experimental techniques and capabilities have expanded significantly in the last decades, but ground state decay intensities remain quite difficult to determine experimentally. We can therefore not expect easy progress for the heaviest mirror nuclei systems (where only ground state branches can be compared), but the increase of the Coulomb effects in these systems makes it interesting to have data here also in order to test our understanding. Moving to systems with higher isospin is not only interesting in itself, it will present more opportunities to find systems where the neutron-rich decay also proceeds to excited states.

Since estimating ground state decays from the corresponding mirror decay has been done often, it may for future use be useful to note from the overall pattern of the current data that this entails a systematic uncertainty in log(ft) that could be as high as 0.05.

The current state of mirror beta transitions, as reflected in Figs 2 to 5, may seem unsatisfactory, but there are good hopes that improvements both on the experimental and theoretical side may help in the near future.
The main challenge is to quantify the theoretical precision in the description of the binding energy effects. Before this has been accomplished we will not be able to use beta decay asymmetries to learn about strong charge symmetry breaking interactions, nor about weak second class currents.

\begin{acknowledgements}
I acknowledge funding from the Independent Research Fund Denmark (9040-00076B) and would like to thank H.O.U. Fynbo and A.S. Jensen for physics discussions and E.A.M. Jensen for assistance with the figures.
\end{acknowledgements}


\end{document}